# Photonic engineering of hybrid metal-organic chromophores

*Mickaël P. Busson, Brice Rolly, Brian Stout, Nicolas Bonod, Jérôme Wenger,\* and Sébastien Bidault\**

Fluorescent probes play a key role in sensing, imaging, and energy harvesting. A common element found among the wide library of existing chromophores is that their photophysical responses have been designed following a molecular engineering approach that tunes the electronic and vibrational eigenstates of the molecules.[1-4] A conceptually different approach is to tailor the molecular photophysical response *via* the external electromagnetic field, which we call photonic engineering. As pioneered by K. H. Drexhage with $Eu^{3+}$ complexes in the vicinity of a mirror,[5] different luminescence properties can be engineered electromagnetically by changing the local density of optical states (LDOS):[6,7] the emission rates, excitation cross-sections and quantum yield of dye molecules, as well as spectroscopic selection rules (electric/magnetic dipolar/quadrupolar transitions) or inter-system crossing rates. Despite the intense recent research on fluorophores coupled to photonic structures,[8-18] one of the main practical features of traditional chromophores, their solubility, remains a challenge. Indeed, a practical implementation of the photonic engineering approach requires combining, in a single hybrid nanostructure, a luminescent molecule and an optical antenna or cavity that confines the electromagnetic field.[19]

In this report, we experimentally demonstrate control of the absorption and emission properties of individual emitters by photonic antennas in suspension. The method results in a new class of water-soluble chromophores with unprecedented photophysical properties, such as short lifetime, low quantum yield but high brightness. We study purified suspensions of 40 nm diameter gold nanoparticles (AuNP) monomers and dimers linked to a single 30 or 50 base pair (bp) DNA double-strand exhibiting a single ATTO647N molecule, as recently demonstrated by our group.[20,21] A combination of fluorescence correlation spectroscopy (FCS), time-correlated fluorescence measurements and spectroscopy fully characterizes these hybrid metal-organic emitters by their absorption cross-sections, fluorescence lifetimes, quantum yields, and translational and rotational diffusion coefficients. These measurements also assess of the high purity of the functionalized AuNP samples with negligible free dyes and weak monomer subpopulations (< 30%) in dimer samples. Compared to isolated chromophores, an organic dye engineered by a AuNP dimer exhibits absorption cross-sections and spontaneous decay rates enhanced by more than one order of magnitude without losing its brightness.

Noble metal nanostructures, that exhibit broad resonant excitations of their valence electrons (plasmon), offer the combination of enhanced local fields and large scattering cross-sections needed to strongly influence the absorption and emission properties of chromophores.[8-13] Nanoparticle dimers are particularly attractive as the electromagnetic field is strongly confined in the nanometre gap that separates the particles.[14] In Purcell factor terms, the strong field confinement compensates the weak quality factor of the plasmon resonance.[22] Using electrophoresis, we synthesize gold nanoparticle dimers linked by a single DNA double-strand, that are stretched by repulsive electrostatic interactions.[20] The interparticle distance is tuned by changing the length of the DNA template, and a single ATTO647N molecule is introduced in the structure.[21] Figure 1a represents the considered nanostructure geometry and Figure 1b provides a typical cryo-electron microscopy (cryo-EM) image of 40 nm AuNPs linked by a single 50 bp DNA strand. We consider five different sample geometries: a reference solution of ATTO647N modified 50 bp DNA strands (*ATTO*), suspensions of single 40 nm diameter AuNPs functionalized with one ATTO647N modified 50 bp or 30 bp DNA strand (*mono50* and *mono30*, respectively) and suspensions of 40 nm AuNP dimers linked by a single ATTO647N modified 50 bp or 30 bp DNA linker (*dim50* and *dim30*).

[∗] M. P. Busson, Dr. S. Bidault
Institut Langevin, ESPCI ParisTech, CNRS UMR 7587, INSERM U979
1 rue Jussieu, 75005 Paris, France
Fax: (+) 33 1 80 96 33 55
E-mail : sebastien.bidault@espci.fr

B. Rolly, Dr. B. Stout, Dr. N. Bonod, Dr. J. Wenger
Institut Fresnel, CNRS, UMR 7249, Aix-Marseille Université, Ecole Centrale Marseille,
Campus de Saint Jérôme, 13397 Marseille, France
E-mail : jerome.wenger@fresnel.fr

The authors thank E. Larquet for the cryo-EM measurement. The research leading to these results has received funding from the Agence Nationale de la Recherche via project ANR 11 BS10 002 02, and the European Research Council under the European Union's Seventh Framework Programme (FP7/2007-2013) / ERC Grant agreement 278242. Work at Institut Langevin is supported by the Laboratoire d'Excellence (LabEx) WIFI.

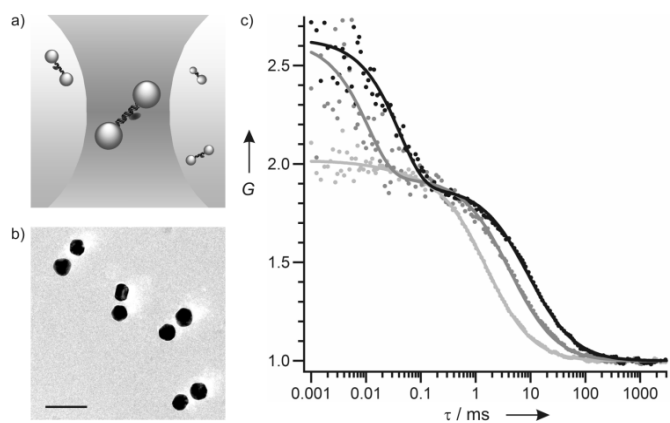

*Figure 1.* (a) DNA templated AuNP dimers as water-soluble hybrid metal-organic chromophores. (b) Cryo-EM image of 50 bp DNA templated 40 nm AuNP dimers (*dim50*, scale bar is 100 nm). (c) Normalized FCS correlation functions and numerical fits for the reference ATTO647N-DNA sample (*ATTO*, light grey), 30 bp monomers (*mono30*, grey) and 30 bp dimers (*dim30*, black).

To demonstrate photonic engineering of water-soluble chromophores, we investigate the sample suspensions with FCS, time-correlated fluorescence measurements and spectroscopy. FCS is a powerful method to accurately quantify the average number of emitters in the confocal detection volume and their average brightness, and infer the sample translational and rotational diffusion properties.[23-25] Fluorescence correlation functions for the *ATTO*, *mono30* and *dim30* samples are shown on Figure 1c for a linearly polarized excitation. These data are normalized to better reveal the changes in the correlation times (raw data are provided in the supporting information SI for all sample geometries). Numerical fitting of the FCS data is based on a standard three dimensional Brownian diffusion model for single species assuming Gaussian excitation and detection mode profiles (see SI for details).[25] Two different characteristic times are readily seen in the FCS data: the long timescale denotes translational diffusion, while the timescale below 100 μs relates to rotational diffusion of the metal-organic chromophores respective to the laser linear polarization.

Table 1 summarizes the translational and rotational diffusion times as well as the translational diffusion coefficients and hydrodynamic radii derived from the diffusion times. Values obtained for the reference *ATTO* sample are consistent with a 50 bp DNA double strand, and gradually increase when adding one or two AuNPs. This behavior further confirms the purity of our monomer and dimer suspensions. The number of free dyes in the monomer samples is estimated below 5%, and negligible for the dimers. Lastly, we infer theoretically the correlation times induced by rotational diffusion for a single sphere and a prolate ellipsoidal particle to mimic the dimer (last column of Table 1).[26] The good agreement with our experimental observations confirms the bumps on the FCS correlation functions below 100 μs relate to rotational diffusion, and not to triplet blinking.

*Table 1.* Diffusion properties of hybrid metal-organic chromophores

| Sample | Diffusion time (ms) | Diffusion coefficient (cm$^2$/s) | Hydro-dynamic radius (nm) | Rotational time (μs) | Rotational time (μs) (theory) |
|---|---|---|---|---|---|
| *ATTO* | 1.53 ± 0.05 | 5.3 10$^{-7}$ | 4.1 | - | - |
| *mono50* | 4.3 ± 0.1 | 1.9 10$^{-7}$ | 11.5 | 11 ± 1 | 11[a] |
| *mono30* | 4.2 ± 0.1 | 1.9 10$^{-7}$ | 11.3 | 11 ± 1 | 11[a] |
| *dim50* | 9.6 ± 0.2 | 8.5 10$^{-8}$ | 25.7 | 36 ± 2 | 36.9[b] |
| *dim30* | 9.2 ± 0.2 | 8.8 10$^{-8}$ | 24.6 | 38 ± 2 | 33.4[c] |

a] 44 nm diameter sphere [b] 40 nm wide, 99 nm long prolate ellipsoid [c] 40 nm wide, 93 nm long prolate ellipsoid

A key parameter for applications is the fluorescence brightness, quantified by the average number of photons detected *per* chromophore. This value is readily obtained by dividing the average fluorescence intensity by the average number of emitters measured by FCS. Figure 2a presents the molecular brightness as a function of the excitation intensity, and Table 2 summarizes the fluorescence enhancement as compared to the reference *ATTO* sample in the excitation regime below 65 μW/μm$^2$. Remarkably, 30 bp AuNP dimers provide 1.35x more signal than the reference *ATTO* sample, while for all other sample geometries the brightness is quenched as compared to the reference. Please note that these values are time-averaged and spatially averaged over all positions and orientations inside the confocal volume; for some given positions and orientations, the fluorescence enhancement can be much higher.

Statistical analysis of cryo-EM images of dimers indicates that 30 bp and 50 bp linkers correspond to 13 ± 2 nm and 17.5 ± 3 nm respectively.[20] Therefore, we only observe enhancement of the fluorescence signal for a single molecule in 40 nm AuNP dimers when the emitter-particle distance falls below 8 nm: enhancement of x1.35 at ~6.5 nm compared to x0.76 at ~9 nm. For monomer samples, the distance dependence of the enhancement factor is reversed with a factor of 3 reduction of ATTO647N fluorescence for a 15 bp DNA spacer (~6.5 nm) compared to a factor of 2 for 25 bp (~9 nm).

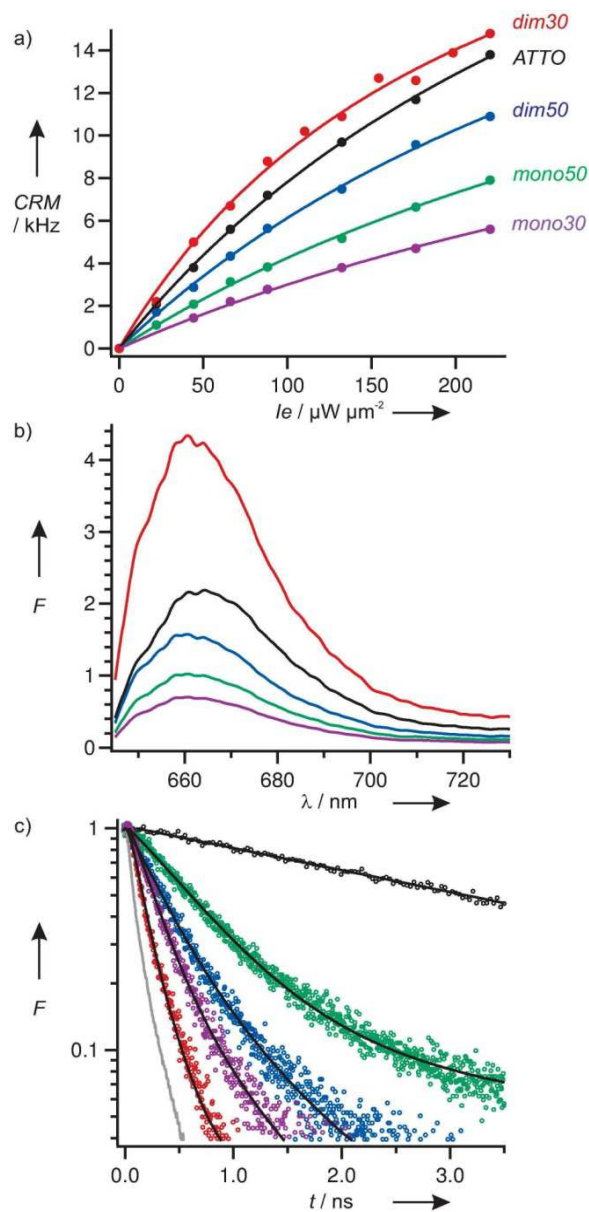

*Figure 2.* Count rate *per* molecule (CRM) as function of the excitation intensity (a), fluorescence spectra normalized *per* emitter (b) and normalized fluorescence decay traces (c) of the reference *ATTO* sample (black), *mono50* (green), *mono30* (purple), *dim50* (blue) and *dim30* (red). Solid lines in (a) correspond to theoretical fits of the saturation curves. Solid black lines in (c) are fitted decay curves. Grey data points in (c) correspond to the instrument response function.

To verify that these signal evolutions are not due to a chemical modification of the luminescent emitter, we measure the fluorescence spectra for the different samples on Figure 2b. All spectra are normalized by the average number of emitters quantified by FCS, and indicate quantitatively the same enhancement and quenching factors as in Figure 2a. Most importantly, all spectra are centered on 670 nm as expected for ATTO647N fluorescence, indicating that the emitter-antenna interaction does not modify the electronic eigenstates of the molecule as expected in a weak coupling regime. ATTO647N absorption and emission peaks are thus strongly red-shifted with respect to the plasmon resonance of DNA templated 40 nm AuNP dimers around 560 nm (see SI).

Figure 2c displays typical time-correlated fluorescence decay traces used to measure the fluorescence lifetimes τ given in Table 2. While the fluorescence decay of the reference and *dim30* samples are fitted with a single exponential curve (convoluted with the instrument response function, see SI), the numerical fitting of the decay traces are consistent with a 2.5 % subpopulation of free dyes for the monomer samples and a 30 % subpopulation of *mono50* in *dim50*. The decay traces thus confirm the sample purities estimated in FCS and are in good agreement with cryo-EM measurements. Table 2 highlights the dramatic influence of AuNPs on the fluorescence lifetimes, which are reduced from 4.3 ns for the *ATTO* reference down to 65 ps for *dim30*. In the case of the dimer samples, these strong modifications are observed with similar luminescence signals compared to a reference *ATTO* sample. It is thus possible to design bright metal-organic emitters with sub-100 ps lifetimes using photonic engineering.

To go further and quantify the chromophore molar extinction coefficient ε and quantum yield ϕ we analyze the saturation curves of Figure 2a by considering a quantized two-level system. Under steady-state conditions, the detected count rate *per* molecule *CRM* is given by [27]

$$CRM = \kappa \phi \frac{\sigma I_e}{1 + I_e / I_{sat}} \quad (1)$$

where κ is the collection efficiency, σ the excitation cross-section, $I_e$ the excitation intensity and $I_{sat}$ the saturation intensity, which can be expressed as a function of the total decay rate $k_{tot} = 1/\tau$ as $I_{sat} = k_{tot} / \sigma$ (neglecting triplet contributions, as confirmed by the FCS curves). Fitting the curves in Figure 2a with Equation (1) allows us to infer the relative changes induced by the AuNPs on the excitation cross-section and quantum yield, respective to the *ATTO* reference. From these values and the tabulated photophysical properties of ATTO647N dyes, we estimate in Table 2 the values of the molecular extinction coefficients ε and quantum yields ϕ for hybrid gold-organic chromophores. Apart for *mono50*, all other samples feature absorption cross-sections and fluorescence quantum yields that are modified by over one order of magnitude. Surprisingly, the *dim30* sample can be as bright as the *ATTO* reference, but with a quantum yield of only 1.6%, thanks to a much higher excitation cross-section that compensates for the quenching ohmic losses in the AuNPs.

*Table 2.* Photophysical properties of hybrid metal-organic emitters

| Sample | Fluorescence enhancement factor [a] | τ (ps) | ε (M$^{-1}$cm$^{-1}$) @ 633nm | ϕ (%) |
| --- | --- | --- | --- | --- |
| ATTO | 1 | 4300 ± 100 | 10$^5$ [b] | 65[c] |
| mono50 | 0.51 ± 0.01 | 615 ± 20 | (3.4 ± 0.4) 10$^5$ | 9.3 ± 1.0 |
| mono30 | 0.35 ± 0.01 | 135 ± 15 | (12 ± 2) 10$^5$ | 1.8 ± 0.3 |
| dim50 | 0.76 ± 0.02 | 185 ± 15 | (11 ± 1.5) 10$^5$ | 4.8 ± 0.7 |
| dim30 | 1.35 ± 0.04 | 65 ± 10 | (57 ± 9) 10$^5$ | 1.6 ± 0.3 |

[a] estimated in the linear, low excitation regime, with respect to *ATTO* [b] extrapolated at 633 nm [c] as provided by the manufacturer for free ATTO647N in water.

To further confirm that our studied samples correspond to purified suspensions of AuNP monomers and dimers linked to one ATTO647N-modified DNA strand, we compare experimental luminescence lifetimes with theoretical estimations of τ. In practice, fluorescence lifetimes are estimated by computing the power dissipated by a point dipole in AuNP monomers and dimers using classical electromagnetic theory.[28,29] Since dye-modified DNA strands feature fixed[30] but uncontrolled molecular orientations, we take an isotropic distribution of the angle between the molecular transition dipole and the particles. The emitter-AuNP distances are set at 6.5 nm and 9.5 nm for the 30 bp and 50 bp linkers, in good agreement with cryo-EM and scattering spectroscopy.[20] With these distances, theoretical values of τ obtained with Mie theory are 574 / 182 ps for the 50 / 30 bp monomers and 198 / 69 ps for the 50 / 30 bp dimers, in excellent agreement with the experimental values given in Table 2.

Classical electromagnetic theory can also be used to highlight how brighter hybrid chromophores can be engineered.[28,29] In particular, the luminescence brightness strongly depends on the orientation of the molecular transition dipole with respect to the dimer axis.[21] If all molecules were oriented along the dimer axis, the theoretical enhancement factors are expected to be increased by a factor of ~3 with maximum ε and minimum τ. This would require rigid chemical linkers between dye molecules and one or two DNA bases. Controlling molecular orientations in hybrid emitter-antenna systems appears essential to design bright ultrafast fluorescent nanostructures.

In conclusion, we have demonstrated that AuNP dimers can be used to electromagnetically engineer the photophysical properties of organic dyes, while preserving their solubility. These hybrid metal-organic emitters were fully characterized by their molar extinction coefficients, fluorescence lifetimes, quantum yields, and translational and rotational diffusion coefficients. In particular, 30 bp DNA-templated dimers feature absorption cross-sections and decay rates enhanced by more than one order of magnitude with respect to isolated dyes while producing high fluorescence count rates. These hybrid chromophores are bright emitters with lifetimes below 100 ps, quantum yields of a few percent and large excitation cross-sections. Experimental results stand in good agreement with calculations based on classical electromagnetic theory which in turn indicates how the emitter brightness can be further optimized in rigid DNA-chromophore complexes with controlled molecular orientations. This work shows that by tuning the emission wavelength of the dye molecule with respect to the plasmon resonance, and by minimizing ohmic losses in the nanostructure, photonic engineering of luminescence will allow the development of unprecedented photophysical properties in hybrid metal-organic chromophores.

*Experimental Section*

Hybrid metal-organic chromophores are synthesized and purified using published procedures.[20,21] In brief, commercial 40 nm AuNPs (BBI, UK) are coated with a negatively charged phosphine ligand (BSPP, Strem Chemicals, USA) then rinsed and concentrated by centrifugation. 30 or 50 bases long trithiolated DNA strands (Fidelity Systems inc., USA) are effectively lengthened by thermal annealing with five consecutive 100 bases long PAGE purified DNA strands (IDT DNA, USA). All DNA sequences are taken from previous published work.[20] The complementary sequences of the 30 and 50 bases long strands are commercially modified with one ATTO647N dye molecule on a central amine-modified base, and also lengthened with five 100 bases long strands. The phosphine coated AuNPs and lengthened DNA strands are left to react overnight in water with a 30x excess of DNA at 30 mM NaCl. The phosphine ligand is replaced by adding an excess of short thiolated methyl-teminated ethylene glycol oligomers (Polypure, Norway) before particles functionalized with 0, 1 or 2 DNA strands are electrophoretically separated in a 1.5% w/w agarose gel. Part of the mono-functionalized AuNPs (with one ATTO647N molecule) is hybridized to unmodified 30 or 50 bases long complementary strands to yield *mono30* and *mono50*. The rest of the samples are hybridized to mono-functionalized AuNPs with the complementary strand (devoid of dye molecule) to produce *dim30* and *dim50*. Hybridization procedures are performed at 55°C to remove the lengthening strands that are hybridized over 15 bp. Monomer and dimer samples are purified twice in electrophoresis (1.5 % w/w agarose) to remove free dye modified DNA strands and monomers from dimer suspensions. The reference *ATTO* sample is obtained by annealing overnight the ATTO647N modified 50 bases long trithiolated strand to a non-thiolated complementary strand at 100 mM NaCl.

Experiments are performed with ~200 pM chromophore concentrations, on a custom developed confocal microscope using a 1.2 NA water immersion objective and linearly polarized 633 nm excitation. FCS measurements are performed under CW illumination by cross-correlating the signal from two avalanche photodiodes with a hardware correlator. Lifetime measurements are performed under picosecond pulsed illumination by a time-correlated single photon counting module displaying 120 ps overall temporal resolution. Full details on the experimental setup, data analysis and raw FCS data are given in the SI.

Theoretical calculations are performed using an in-house Generalized Mie Theory code.[29] The emission wavelength is set at 670 nm and the particle diameter at 40 nm. The refractive index of the surrounding medium is set at 1.335, while the dielectric constant of gold is tabulated from published data.[31] Theoretical fluorescence lifetime estimations are performed up to a multipolar order of 30 and by taking into account that the brightness depends on the molecular transition dipole orientation.

**Experimental setup**

The experimental setup combines FCS and time-correlated lifetime measurements facilities on the same confocal microscope equipped with a 40x 1.2NA water immersion objective (Zeiss C-Apochromat). For FCS measurements, the excitation source is a linearly polarized CW He-Ne laser operating at 633 nm. To ensure that at least 0.2 fluorescent emitters are detected in the confocal volume while the typical concentrations are about 200 pM, we underfill the microscope objective back aperture (4 mm instead of 8.9 mm) and use a 100 µm confocal pinhole conjugated to the object plane. FCS experiments on Alexa Fluor solutions calibrate the confocal volume to 4.5 fL and the transversal waist to $w_{xy}$ = 570 nm. After the confocal pinhole, the detection is performed by two avalanche photodiodes (PicoQuant MPD-5CTC). To separate the fluorescence light from the epi-reflected laser and elastically scattered light, we use a dichroic mirror (Omega Filters 650DRLP), a long pass filter (Omega Filters 640AELP) and a 670 ± 20 nm fluorescence bandpass filter (Omega Filters 670DF40) in front of each photodiode. For FCS, the fluorescence intensity temporal fluctuations are analyzed by cross-correlating the signal from the photodiodes with a ALV6000 hardware correlator. Each individual FCS measurement is obtained by averaging at least 20 runs of 10 s duration each.

For lifetime measurements, the excitation source is switched to a picosecond laser diode operating at 636 nm (PicoQuant LDH-P-635, repetition rate 80 MHz). A single-mode optical fiber (Thorlabs P3-630A-FC-5) ensures a perfect spatial overlap between the laser beams and guarantees the same alignment and confocal volume for FCS and lifetime measurements. The fast timing output of the photodiode is coupled to a time-correlated single photon counting (TCSPC) module (PicoQuant PicoHarp300). The temporal resolution of our setup for fluorescence lifetime measurements is 120 ps FWHM. Analysis of the instrument response function (IRF) reveals a double exponential decay: IRF(t) = $A_1$ exp(-$k_1$ t) + $A_2$ exp(-$k_2$ t) with $A_1$ = 0.65, $A_2$ = 0.35, $k_1$ = 17.24 $10^9$ s$^{-1}$ and $k_2$ = 4.59 $10^9$ s$^{-1}$. The output signal S(t) of the TCSPC card corresponds to the convolution of the system IRF with the fluorescence decay, which is assumed to be mono-exponential. Convoluting a single-exponential fluorescence decay with a double exponential IRF results in a triple exponential, which is used to fit the TCSPC data. In the data numerical fitting, the fluorescence lifetime is the only free varying parameter while $A_1$, $A_2$, $k_1$ and $k_2$ are fixed parameters set by the IRF analysis. Figure S1 diplays the result of the fit for the 30bp dimer case, along with the IRF. Note the linear scale used here for the vertical axis, which better highlights the quality of the data fitting.

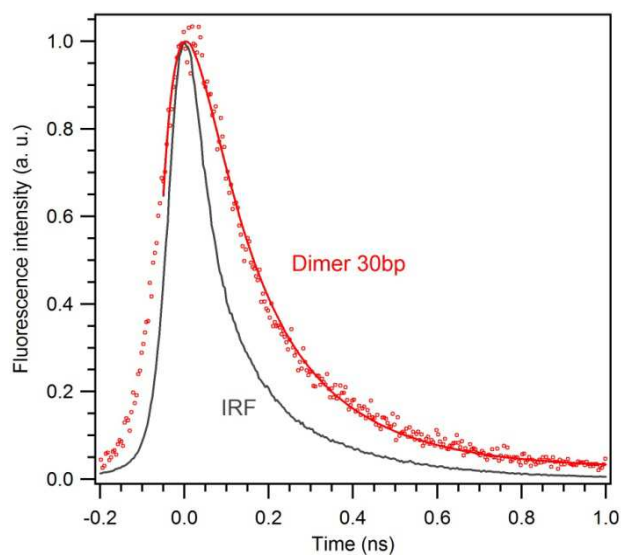

*Figure S1.* Fluorescence decay curve for the 30bp dimer case, the solid red line is a fit taking into account the convolution by the IRF, and results in a fluorescence lifetime of 65 ps.

**Raw FCS data**

Figures S2 to S6 present the non-normalized data corresponding to Figure 1c and Table 1. Each figure also presents the residuals of the FCS data fit. The excitation power is 65 µW here, and the background noise is 0.13 kHz (corresponding to 0.5% of the time-averaged fluorescence intensity for the ATTO sample and between 3% and 10% for the monomer and dimer samples that feature lower concentrations). For clarity, only the first 100 s of the time trace are displayed (binning time for display 40 ms). The lower part of each figure shows the residuals of the fit.

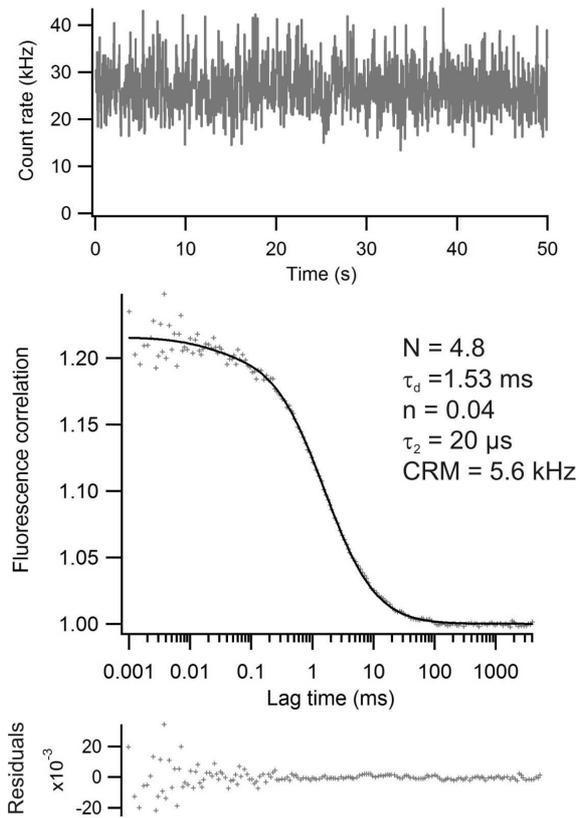

*Figure S2.* FCS data for the reference ATTO647N-DNA sample.

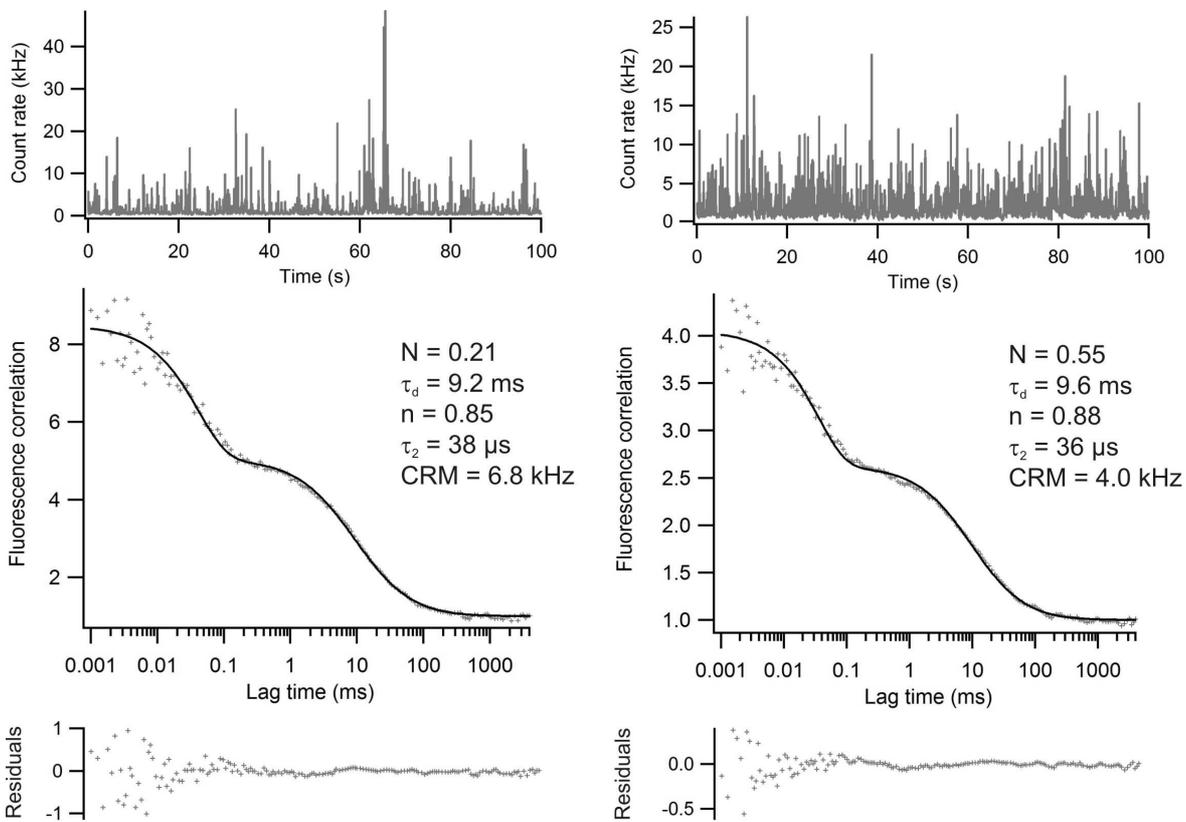

*Figure S3.* FCS data for the 30bp dimer sample.   *Figure S4.* FCS data for the 50bp dimer sample.

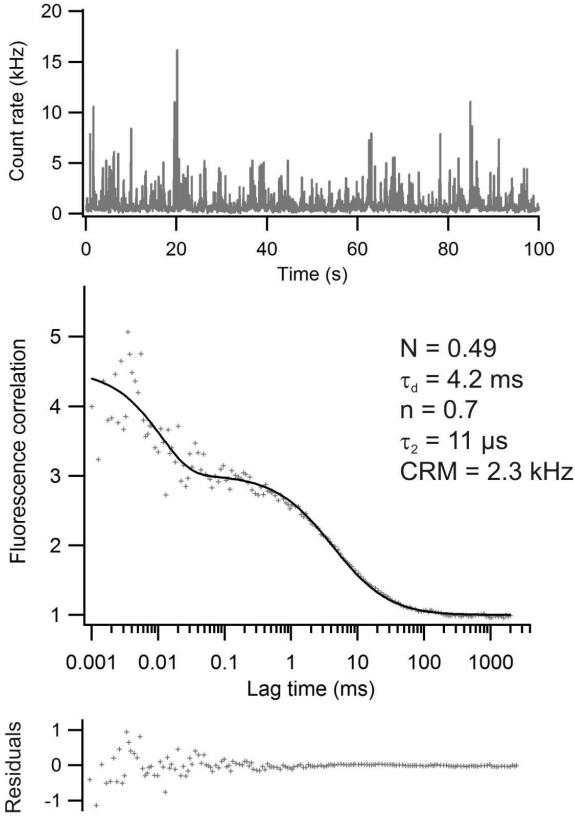 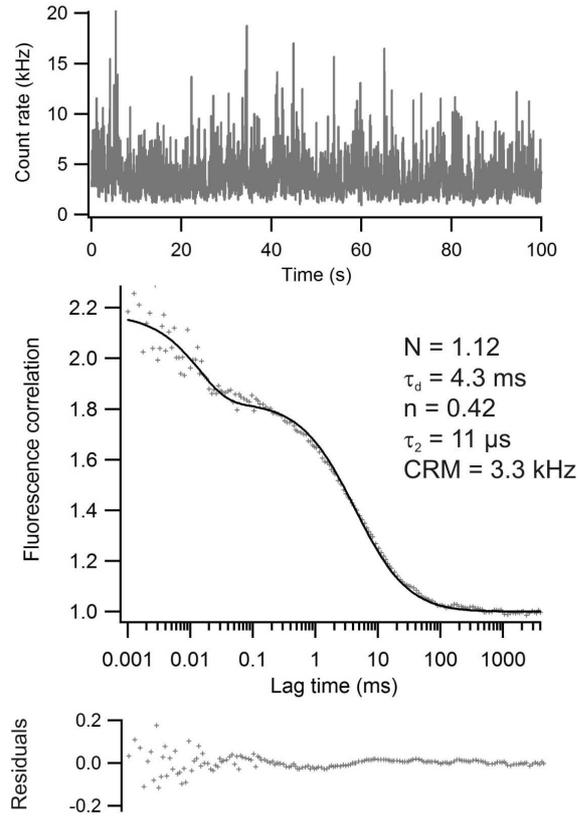

*Figure S5*. FCS data for 30bp monomer sample.   *Figure S6*. FCS data for 50bp monomer sample.

**FCS data analysis**

The analysis of the FCS data relies on a numerical fit based on a three dimensional Brownian diffusion model assuming Gaussian excitation and detection profiles:

$$g^{(2)}(\tau) = 1 + \frac{1}{N}\left(1 - \frac{B}{F}\right)^2 \left(1 + n\exp\left(-\frac{\tau}{\tau_2}\right)\right) \frac{1}{\left(1 + \frac{\tau}{\tau_d}\right)\sqrt{1 + s^2 \frac{\tau}{\tau_d}}} \quad (S1)$$

$N$ is the total number of emitters, $F$ the total fluorescence intensity, $B$ the background noise, $n$ the rotational correlation amplitude, $\tau_2$ the rotational diffusion time, $\tau_d$ the transverse translational diffusion time and $s$ the ratio of transversal to axial dimensions of the analysis volume, which we set to $s = 0.2$. A simplified version of the contribution accounting for rotational diffusion is used here as the excitation is linearly polarized and the detection collects all orientations [1-3]. The rotational diffusion time $\tau_2$ relates to the rotational coefficient $\Theta$ by $\tau_2 = 1/6\Theta$, and the translational diffusion times relates to the diffusion coefficient $D$ by $\tau_d = w_{xy}^2 / 4D$. Lastly, the hydrodynamic radius $R_h$ is given by the Stokes-Einstein equation:

$$D = \frac{k_B T}{6\pi \eta R_h} \tag{S2}$$

where $k_B$ is Boltzmann's constant, $T$ the absolute temperature, and $\eta$ the solvent's viscosity.

**Spectral analysis**

To investigate spectral modifications of the emission spectrum of ATTO647N, the fluorescence beam after the confocal pinhole is sent to a spectrograph (Jobin-Yvon SPEX 270M) equipped with a nitrogen-cooled CCD detector. The raw spectrum is then normalized by the number of molecules as measured by FCS. This yields the average fluorescence spectra *per single molecule* displayed in Figure 2b. It is then straightforward to compare the spectra computed back to *per molecule*, and deduce the fluorescence enhancement factor for each emission wavelength, which is the ratio of the spectrum recorded for a photonic-engineered emitter by the reference ATTO647N-DNA spectrum (Figure S7). Within the range 650-690 nm, the apparent value of the fluorescence enhancement slightly decreases with the wavelength, showing a higher enhancement for blue-shifted wavelengths. The emission spectra for the different samples show qualitatively the same behavior, which is explained by the blue-shifted position of the main plasmon resonance[4] respective to the peak ATTO647N emission wavelength.[5]

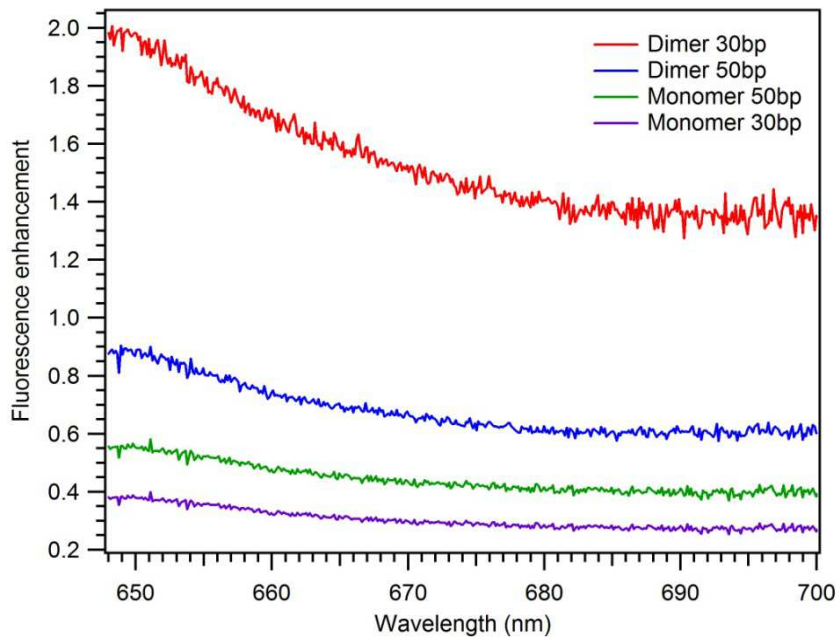

***Figure S7.*** Fluorescence enhancement factor as a function of the emission wavelength, as compared to the reference ATTO647N-DNA sample.

**Scattering spectra of hybrid emitters**

Scattering spectra of monomers and 30 bp and 50 bp dimers are measured on surface immobilized single nanostructures using a transmission darkfield microscope as described elsewhere.[4] Figure S8 shows typical scattering spectra of a 40 nm AuNP and 50 bp and 30 bp dimers. The single AuNP features the expected gold particle plasmon resonance at 542 nm while dimers exhibit a red-shifted longitudinal plasmon mode with increased scattering efficiency. In this example, there is a weak red-shift from 557 nm to 563 nm when shortening the DNA linker by 20 base pairs. A 5 nm average red-shift was observed in a statistical analysis of 36 nm AuNP dimers with the same DNA spacers.[4] In all cases, the plasmon resonance is strongly blue-shifted with respect to the ATTO647N emission wavelength.

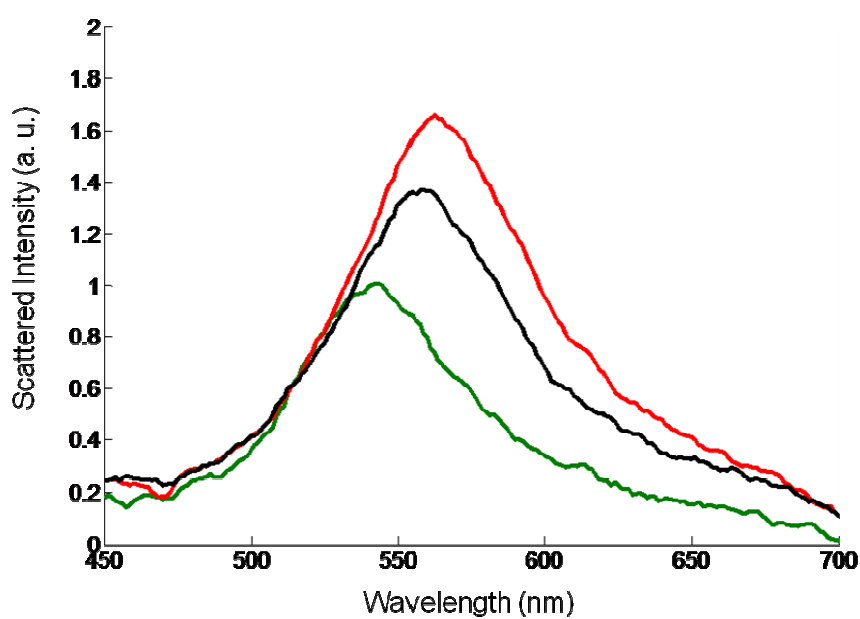

*Figure S8.* Scattering spectra of a single 40 nm AuNP (green) and 50 bp (black) and 30 bp (red) dimers. The arbitrary units correspond to the same EMCCD integration time and are corrected for background scattering and the spectral shape of the white light illumination.